\documentclass[Afour,sagev,times]{sagej}

\usepackage{moreverb}
\usepackage{url}
\usepackage[colorlinks,bookmarksopen,bookmarksnumbered,citecolor=red,urlcolor=red]{hyperref}
\newcommand\BibTeX{{\rmfamily B\kern-.05em \textsc{i\kern-.025em b}\kern-.08em
T\kern-.1667em\lower.7ex\hbox{E}\kern-.125emX}}

\newcommand{\bmath}[1]{{\mbox{\boldmath $#1$}}}
\newtheorem{Def}{Definition}
\usepackage{kbordermatrix}
\usepackage{natbib}
\usepackage{color}
\usepackage{colortbl}
\usepackage{listings}
\newtheorem{example}{Example}

\begin{document}
\setlength{\abovedisplayskip}{0pt}
\setlength{\belowdisplayskip}{2pt}
\runninghead{Boolean Monte Carlo method}

\title{A method obtaining trends of boolean operators and constructing 2 by 2 contingency table in multivariate case: Boolean Monte Carlo method}

\author{Takuma Usuzaki\affilnum{1}, Minoru Shimoyama\affilnum{1}, Shuji Chiba\affilnum{2}, Naoko Mori\affilnum{3} and Shunji Mugikura\affilnum{3}}

\affiliation{\affilnum{1}Tohoku University School of Medicine, Sendai, Japan\\
\affilnum{2}Dokkyo Medical University, Tochigi, Japan\\
\affilnum{3}Department of Diagnostic Radiology, Tohoku University Graduate School of Medicine, Sendai, Japan
}

\corrauth{Takum Usuzaki,
Tohoku University School of Medicine,
Seiryo Machi 1-1,
Aoba-ku,
Sendai,
Miyagi, 
980-8574, Japan
}

\email{takuma.usuzaki.p6@dc.tohoku.ac.jp}

\begin{abstract}
Accuracy of medical test and diagnosis are often discussed by 2 by 2 contingency tables. However, it is difficult to apply a 2 by 2 contingency table to multivariate cases because the number of possible categories increases exponentially. The aims of this study is introducing a method by which we can obtain trends of boolean operators and construct a 2 by 2 contingency tables to multivariate cases. In this method, we randomly assigned boolean operators between binary variables and focused on frequencies of boolean operators which could explain outcomes correctly.  We determined trends of boolean operators by chi-square test. As an application, we performed this method for a dataset which included patient's age, body mass index (BMI), mean arterial pressure (MAP) and the quantitative measure of diabetes progression ($Y$). We set cutoff to each variable and considered these as binary variables. Interactions of age, BMI and MAP were determined as age $and$ BMI $and$ MAP ($p<0.0001$ for both operators) in estimating $Y$ and sensitivity, specificity, positive and negative predicting value were 0.50, 0.85. 0.59, 0.80 respectively. We may be able to detemine trends of boolean operators and construct a 2 by 2 contingency table in multivariate situation by this method.
\end{abstract}


\maketitle

\section{Introduction}
Boolean algebra is a field of mathematics.\cite{givant_introduction_2009} It has been applied to other fields, e.g., circuit of computer science, cryptography, and medicine.\cite{wu_boolean_2016}
In most these applications, boolean function $f:\{0,1\}^n\rightarrow\{0,1\}$ is constructed by ``$not$,'' ``$and$,'' ``$or$,'' and “parentheses.” In medicine, binary valuese 0 and 1 are often used to represent a result of medical examination and medical test (negative or positive)  and to construct a contingency table.\cite{YARDIMCI20091029} Contingency tables enable us to discuss such as preventive effect, medical test accuracy and treatment effect.\cite{DAE66884, 10.1371/journal.pone.0224460} In case variables are continuous, we can use binary values by setting cutoff values. However, it is difficult to apply a contingency table to a multivariate situation. When we consider $N$ binary variables, the number of possible categories is $2^N$. We cannot get enough sample size in each category for large $N$ and obtain sufficient statistical power. Under insufficient statistical power, it is difficult to determine interactions of variables. A method modeling interactions should be considered because a patient has multiple symptoms or medical test results in clinical practice.\cite{10.1371/journal.pmed.0030208} To resolve these difficulties, we propose a method in which boolean operators are randomly assigned and frequencies of boolean operators are tested by chi-square test. By these processes, we can obtain the trends of operators and construct 2 by 2 contingency table in multivariate case. The aim of this paper is introducing a method called the Boolean Monte Carlo method (BMCM) by which we can obtain trends of boolean operators and applying this method for real datasets.

\section{Methods}
\subsection{Definitions of some terms and functions}

$0_{B}$ and $1_{B}$ denote binary values. $\wedge$  and $\vee$ denote ``$and$'' and ``$or$'' respectively. $0_{B}$, $1_{B}$, $\wedge$ and $\vee$ satisfy calculation rules shown in \textbf{\textsf{Table}} \textsf{\textbf{\ref{tb:tab0}}}. When $\bmath{A}$ is a $m\times n$ matrix, $a_{ij}, a_{i\bullet}$ and $a_{\bullet j}$ denote element, $i$-th row and $j$-th column of $\bmath{A}$ respectively $(i,j\in\mathbb{N}, 1\leq i\leq m$ and $1\leq j\leq n)$. When we need to avoid confusion, we denote $a_{ij}$ as $a_{i, j}$.

\begin{table}
\small\sf\centering
\caption{Calculation rules.}
\begin{tabular}{cc}
\toprule
$\wedge (and)$& $\vee (or)$ \\ \midrule
$0_{B}\wedge0_{B}=0$& $0_{B}\vee0_{B}=0$\\
$0_{B}\wedge1_{B}=0$&  $0_{B}\vee1_{B}=1$\\
$1_{B}\wedge0_{B}=0$&  $1_{B}\vee0_{B}=1$\\ 
$1_{B}\wedge1_{B}=1$&  $1_{B}\vee1_{B}=1$\\ 
\bottomrule
\label{tb:tab0}
\end{tabular}
\end{table}

\begin{Def}[\textbf{Boolean matrix and operator matrix}]
Let $\bmath{A}$ be a matrix. When all elements of  $\bmath{A}$ are $0_{B}$ or $1_{B}$, we call  $\bmath{A}$ ``boolean matrix." When all elements of  $\bmath{A}$ are $\wedge$ or $\vee$, we call  $\bmath{A}$ ``operator matrix."
\end{Def}
\begin{example}
\normalfont
Matrices \bmath{A} and \bmath{B} are boolean matrix and operator matrix respectively;
\begin{eqnarray*}
\begin{split}
  \bmath{A} =\kbordermatrix{
    &\textsf{1}&\textsf{2}&\textsf{3}&\textsf{4} \\
    \textsf{1}&0_B&0_B&0_B&0_B \\
    \textsf{2}&1_B&1_B&1_B&1_B \\
    }
\end{split}
\begin{split}
  ,\bmath{B} =\kbordermatrix{
    &\textsf{1}&\textsf{2} \\
    \textsf{1}&\wedge&\wedge \\
    \textsf{2}&\vee &\vee \\
    }.
\end{split}
\end{eqnarray*}
\end{example}

\begin{Def}[\textbf{1-row and 0-row}]
$i,j,m,n\in\mathbb{N}, 1\leq i\leq m, 1\leq j\leq n$ and $n>1$. Let $\bmath{A}$ be a $m\times n$ boolean matrix. When $a_{ij}=1$ for all $j (1\leq j\leq n-1)$ we call $a_{i\bullet}$ ``1-row".  When $a_{ij}=0$ for all $j (1\leq j\leq n-1)$ we call $a_{i\bullet}$ ``0-row".  
\end{Def}
\begin{example}\normalfont
In a $4\times 4$ boolean matrix \bmath{A} below, $a_{1\bullet}$ and  $a_{4\bullet}$ are 0-row and 1-row respectively;
\begin{align*}
\begin{split}
  \bmath{A} =\kbordermatrix{
    &\textsf{1}&\textsf{2}&\textsf{3}&\textsf{4} \\
    \textsf{1}&0_B&0_B&0_B&0_B \\
    \textsf{2}&0_B&0_B&1_B&1_B \\
    \textsf{3}&0_B&1_B&1_B&1_B \\
    \textsf{4}&1_B&1_B&1_B&0_B \\
    }.
\end{split}
\end{align*}
\end{example}

\begin{Def}[\textbf{Null row}]
$i,j,m,n\in\mathbb{N}, 1\leq i\leq m, 1\leq j\leq n$ and $ n>1$. Let $\bmath{A}$ be a $m\times n$ boolean matrix. When (i) $a_{i\bullet}$ is 1-row and $a_{in}=1$, (ii) $a_{i\bullet}$ is 1-row and $a_{in}=0$, (iii) $a_{i\bullet}$ is 0-row and $a_{in}=1$ and (iv) $a_{i\bullet}$ is 0-row and $a_{in}=0$, we define  $a_{i\bullet}$ ``null row." 
\end{Def}
\begin{example}\normalfont
In a $4\times 4$ boolean matrix \bmath{A} below, $a_{1\bullet}$ and $a_{4\bullet}$ are null rows;
\begin{align*}
\begin{split}
  \bmath{A} =\kbordermatrix{
    &\textsf{1}&\textsf{2}&\textsf{3}&\textsf{4} \\
    \textsf{1}&0_B&0_B&0_B&0_B \\
    \textsf{2}&0_B&0_B&1_B&1_B \\
    \textsf{3}&0_B&1_B&1_B&1_B \\
    \textsf{4}&1_B&1_B&1_B&0_B \\
    }.
\end{split}
\end{align*}
\end{example}

\begin{Def}[\textbf{The number of null rows}]
Let $\bmath{A}$ be a boolean matrix.{ \rm\#$_{\rm null}$(\bmath{A}, i), \#$_{\rm null}$(\bmath{A}, ii), \#$_{\rm null}$(\bmath{A}, iii)} and { \rm\#$_{\rm null}$(\bmath{A}, iv)} denote the number of rows which belong to (i), (ii) , (iii) and (iv) respectively in $\bmath{A}$.
\end{Def}
\begin{example}\normalfont
In a $4\times 4$ boolean matrix \bmath{A} below
\begin{align*}
\begin{split}
  \bmath{A} =\kbordermatrix{
    &\textsf{1}&\textsf{2}&\textsf{3}&\textsf{4} \\
    \textsf{1}&0_B&0_B&0_B&0_B \\
    \textsf{2}&0_B&0_B&1_B&1_B \\
    \textsf{3}&0_B&1_B&1_B&1_B \\
    \textsf{4}&1_B&1_B&1_B&0_B \\
    },
\end{split}
\end{align*}
\#$_{\rm null}(\bmath{A}, {\rm i})=0$, \#$_{\rm null}(\bmath{A}, {\rm  ii})=1$, \#$_{\rm null}(\bmath{A}, {\rm iii})=0$ and \#$_{\rm null}(\bmath{A},  {\rm iv})=1$.
\end{example}

\begin{Def}[\textbf{A boolean matrix which dose not contain any null row}]
$i, j, k, l, m, n, N\in\mathbb{N}, 1\leq i\leq m, 1\leq j, l \leq n, 1\leq N\leq m$ and $1\leq k\leq N$. Let $\bmath{A}$ be a $m\times n$ boolean matrix of which element is denoted by $a_{ij}$. $i_1, i_2, \cdots i_{N-1}, i_N\in\mathbb{N}$ and $1\leq i_1< i_2<\cdots < i_{N-1}<i_N\leq m$. When $a_{i_1\bullet}, a_{i_2\bullet}, \cdots a_{i_{N-1}\bullet}, a_{i_N\bullet}$ are not null row, we define $a^-_{kl}$ which is the element of boolean matrix of $\bmath{A}^-$ as
$a^-_{k\bullet}=a_{i_k\bullet} (1\leq k\leq N)$.
\end{Def}
\begin{example}\normalfont
A $4\times 4$ boolean matrix $\bmath{A}$ gives $\bmath{A}^-$ as follows;
\begin{align*}
\begin{split}
  \bmath{A} =\kbordermatrix{
    &\textsf{1}&\textsf{2}&\textsf{3}&\textsf{4} \\
    \textsf{1}&0_B&0_B&0_B&0_B \\
    \textsf{2}&0_B&0_B&1_B&1_B \\
    \textsf{3}&0_B&1_B&1_B&1_B \\
    \textsf{4}&1_B&1_B&1_B&0_B \\
    },
\end{split}
\end{align*}
\begin{align*}
\begin{split}
  \bmath{A}^- =\kbordermatrix{
    &\textsf{1}&\textsf{2}&\textsf{3}&\textsf{4} \\
    \textsf{1}&0_B&0_B&1_B&1_B \\
    \textsf{2}&0_B&1_B&1_B&1_B \\
    }.
\end{split}
\end{align*}
\end{example}
\begin{Def}[\textbf{Dataset matrix}]
 $i,j,m,l,n\in\mathbb{N}, n>1, 1\leq i\leq m\times l$ and $1\leq j\leq 2n-2$. $\bmath{A}$ and $\bmath{B}$ are $m\times n$ boolean matrix and $l\times n-2$ operator matrix respectively. We define operation denoted by $\otimes$ which gives a $(m\times l)\times 2n-2$ matrix from $\bmath{A}$ and $\bmath{B}$. Let $\bmath{C}$ be a $(m\times l)\times 2n-2$ matrix and $\bmath{C}=\bmath{A}\otimes\bmath{B}$. We define  the element $c_{ij}$ which is the element of $\bmath{C}$ as  
\[
  c_{ij}=\begin{cases}
	a_{q(i-1)+1,\frac{j+1}{2}} &\textrm{when } j \textrm{ is odd and }  1\leq j\leq 2n-3\\
	b_{r(i),\frac{j}{2}} &\textrm{when }  j \textrm{ is even and } 1\leq j\leq 2n-3\\
	a_{q(i-1)+1,n} & \textrm{when }  j=2n-2.
  \end{cases}
\]
$q(i)$ and $r(i)$ are quotient and remainder  respectively when $i$ divided by $l$. When $r(i)=0$, we define $r(i)=l$. We call $(m\times l)\times 2n-2$ matrix $\bmath{C}$ which is generated from boolean matrix and operator matrix ``dataset matrix." 
\end{Def}
\begin{example}\normalfont
When $\bmath{A}^-$ and $\bmath{B}$ are $2\times 4$ boolean matrix and $2\times 2$ operator matrix respectively below. A $4\times 6$ dataset matrix $\bmath{C}=\bmath{A}^-\otimes\bmath{B}$ is caluclated as follows;
\begin{align*}
\begin{split}
  \bmath{A}^- =\kbordermatrix{
    &\textsf{1}&\textsf{2}&\textsf{3}&\textsf{4} \\
    \textsf{1}&0_B&0_B&1_B&1_B \\
    \textsf{2}&0_B&1_B&1_B&1_B \\
    },
\end{split}
\begin{split}
\bmath{B} =\kbordermatrix{
    &\textsf{1}&\textsf{2} \\
    \textsf{1}&\wedge&\wedge \\
    \textsf{2}&\vee &\vee \\
    },
\end{split}  
\end{align*}
\begin{align*}
\begin{split}
 \bmath{C}= \bmath{A}^-\otimes\bmath{B}  =\kbordermatrix{
    &\textsf{1}&\textsf{2}&\textsf{3}&\textsf{4}&\textsf{5}&\textsf{6} \\
    \textsf{1}&0_B&\wedge&0_B&\wedge&1_B&1_B \\
    \textsf{2}&0_B&\vee&0_B&\vee&1_B&1_B \\
    \textsf{3}&0_B&\wedge&1_B&\wedge&1_B&1_B \\
    \textsf{4}&0_B&\vee&1_B&\vee&1_B&1_B \\
    }.
\end{split}
\end{align*}
\end{example}
\begin{Def}[\textbf{0-faithful row, 1-faithful row, 0-unfaithful row and 1-unfaithful row}]
$i,m,n\in\mathbb{N}, 1\leq i\leq m$ and $n>1$. Let $\bmath{A}$ be a $m\times 2n-2$ dataset matrix. Let us consider calculations $a_{i,1}a_{i,2}\cdots a_{i,2n-4}a_{i,2n-3}$. When $a_{i,1}a_{i,2}\cdots a_{i,2n-4}a_{i,2n-3}=a_{i,2n-2}$ and $a_{i,2n-2}=0_B (1_B)$, we call  $a_{i\bullet}$ ``0(1)-faithful row." 

\noindent When $a_{i,1}a_{i,2}\cdots a_{i,2n-4}a_{i,2n-3}\neq a_{i,2n-2}$ and $a_{i,2n-2}=0_B (1_B)$, we call  $a_{i\bullet}$ ``0(1)-unfaithful row."  \#$_{\rm 0f}$($\bmath{A}$), { \rm\#$_{\rm 1f}$($\bmath{A}$), \#$_{\rm 0u}$($\bmath{A}$)} and { \rm\#$_{\rm 1u}$($\bmath{A}$)} denote the number of 0-faithful row, 1-faithful row, 0-unfaithful row and 1-faithful row respectively in $\bmath{A}$.
\end{Def}
\begin{example}\normalfont
When $\bmath{A}$ is a $4\times 6$ dataset matrix below
\begin{align*}
\begin{split}
\bmath{A} =\kbordermatrix{
    &\textsf{1}&\textsf{2}&\textsf{3}&\textsf{4}&\textsf{5}&\textsf{6} \\
    \textsf{1}&0_B&\wedge&0_B&\wedge&1_B&1_B \\
    \textsf{2}&0_B&\vee&0_B&\vee&1_B&1_B \\
    \textsf{3}&0_B&\wedge&1_B&\wedge&1_B&1_B \\
    \textsf{4}&0_B&\vee&1_B&\vee&1_B&1_B \\
    }, 
\end{split}
\end{align*}
$a_{11}a_{12}a_{13}a_{14}a_{15}=0_B\wedge0_B\wedge1_B=0_B\neq a_{16}$ and $a_{1\bullet}$ is 1-unfaithful row.
$a_{21}a_{22}a_{23}a_{24}a_{25}=0_B\vee0_B\vee1_B=1_B = a_{26}$ and $a_{2\bullet}$ is 1-faithful row. Similarly, $a_{3\bullet}$ is 1-unfaithful row and $a_{4\bullet}$ is 1-faithful row.  In this example, \#$_{\rm 0f}(\bmath{A})=0$, \#$_{\rm 1f}(\bmath{A})=2$, \#$_{\rm 0u}(\bmath{A})=0$ and \#$_{\rm 1u}(\bmath{A})=2$.
\end{example}
\begin{Def}[\textbf{Faithful dataset matrix}]
$i, j, k, l, m, n, N\in\mathbb{N}, 1\leq i\leq m, 1\leq j, l \leq n, 1\leq N\leq m$ and $1\leq k\leq N$. Let $\bmath{A}$ be a $m\times n$ dataset matrix of which element is denoted by $a_{ij}$. $i_1, i_2, \cdots i_{N-1}, i_N\in\mathbb{N}$ and $1\leq i_1< i_2<\cdots < i_{N-1}<i_N\leq m$. When $a_{i_1\bullet}, a_{i_2\bullet}, \cdots a_{i_{N-1}\bullet}, a_{i_N\bullet}$ are 0-faithful or 1-faithful rows. We define $a^{f}_{kl}$ which is the element of boolean matrix of $\bmath{A}^{f}$ as
$a^{f}_{k\bullet}=a_{i_k\bullet} (1\leq k\leq N)$. We call $\bmath{A}^{f}$ ``faithful dataset matrix."
\end{Def}
\begin{example}\normalfont
When $\bmath{A}$ is a $4\times 6$ dataset matrix
\begin{align*}
\begin{split}
\bmath{A} =\kbordermatrix{
    &\textsf{1}&\textsf{2}&\textsf{3}&\textsf{4}&\textsf{5}&\textsf{6} \\
    \textsf{1}&0_B&\wedge&0_B&\wedge&1_B&1_B \\
    \textsf{2}&0_B&\vee&0_B&\vee&1_B&1_B \\
    \textsf{3}&0_B&\wedge&1_B&\wedge&1_B&1_B \\
    \textsf{4}&0_B&\vee&1_B&\vee&1_B&1_B \\
    }, 
\end{split}
\end{align*}
a faithful dataset matrix $\bmath{A}^f$ is
\begin{align*}
\begin{split}
\bmath{A}^f =\kbordermatrix{
    &\textsf{1}&\textsf{2}&\textsf{3}&\textsf{4}&\textsf{5}&\textsf{6} \\
    \textsf{1}&0_B&\vee&0_B&\vee&1_B&1_B \\
    \textsf{2}&0_B&\vee&1_B&\vee&1_B&1_B \\
    }.
\end{split}
\end{align*}
\end{example}
\begin{Def}Let $O$ be a boolean operator. 

We define two functions $f_{\wedge}, f_{\vee}:\{\wedge, \vee\}\rightarrow\{0,1\}\in\mathbb{N}$ as
\[
  f_{\wedge}(O) = \begin{cases}
    1 & (O=\wedge) \\
    0 & (O=\vee)
  \end{cases}
\]
\[
  f_{\vee}(O) = \begin{cases}
    0 & (O=\wedge) \\
    1 & (O=\vee).
  \end{cases}
\]
\end{Def}
\begin{example}\normalfont
When $\bmath{A}$ is a $2\times 6$ faithful dataset matrix
\begin{align*}
\begin{split}
\bmath{A} &=\kbordermatrix{
    &\textsf{1}&\textsf{2}&\textsf{3}&\textsf{4}&\textsf{5}&\textsf{6} \\
    \textsf{1}&0_B&\vee&0_B&\vee&1_B&1_B \\
    \textsf{2}&0_B&\vee&1_B&\vee&1_B&1_B \\
    },
\end{split} \\
&\sum_{i=1}^{2}f_{\wedge}(a_{i2})=0,\;  \sum_{i=1}^{2}f_{\wedge}(a_{i4})=0,  \\
&\sum_{i=1}^{2}f_{\vee}(a_{i2})=2\; \textrm{and}\; \sum_{i=1}^{2}f_{\wedge}(a_{i4})=2.
\end{align*}

\end{example}
A source code for Python language which implement some calculations for matrices is shown in \textbf{Supplemental material S.1}.

\subsection{Explanation of the method}

This method was composed of three steps. From here on, $i,j,m,n\in\mathbb{N}$ and $1\leq i\leq m, 1\leq j\leq n, n>2$, let $\bmath{A}$ be a $m\times n$ boolean matrix and let $\bmath{B}$ be a $2^{n-2}\times n-2$ operator matrix of which rows are different each other. By this definition, $\bmath{B}$ contains the all combinations of $\wedge$ and $\vee$ in its rows. In an application to a dataset, we consider $a_{\bullet j} (1\leq j\leq n-1)$ as explanatory variables and $a_{\bullet n}$ as objective variables.

\subsection*{First step}

For $\bmath{A}$, we create a 2 by 2 contingency table as shown in \textsf{\textbf{Table}} \textsf{\textbf{\ref{tb:tab1}}} and perform a chi-square test (Fisher’s exact test).  When chi-square test for this table is significant, we define $\bmath{A}$ has "rough trend". 

\begin{table}
\small\sf\centering
\caption{Table created in the first step.}
\begin{tabular}{ccc} 
\toprule
Property of row/column & $a_{in}=1$ & $a_{in}=0$ \\ \midrule
1-row & \#$_{\rm null}$(\bmath{A}, i) & \#$_{\rm null}$(\bmath{A}, ii) \\
0-row& \#$_{\rm null}$(\bmath{A}, iii) & \#$_{\rm null}$(\bmath{A}, iv) \\ 
\bottomrule
\label{tb:tab1}
\end{tabular}
\end{table}

We define 4 kinds of null for a real dataset as follows: i) all explanatory variables are $1_B$ and the objective variable is $1_B$, ii) all explanatory variables are $0_B$ and the objective variable is $0_B$, iii) all explanatory variables are $1_B$ and the objective variable is $0_B $and iv) all explanatory variables are $0_B$ and the objective variable is $1_B$. These definitions are required because results of boolean operations for null rows are trivial and operators need not to be determined. For example, when all explanatory variables are $0_B$ and the  objective variable is $1_B$ in a row, we cannot obtain boolean operation which explain the objective variable whether $\wedge$ or $\vee$ was used.

\subsection*{Second step}

In the second step, we consider a matrix $\bmath{X}=(\bmath{A}^-\otimes\bmath{B})^{f}$. We assume that $\bmath{X}=(\bmath{A}^-\otimes\bmath{B})^{f}$ is a $N\times 2n-2 (N\in\mathbb{N}, 1\leq N\leq m\times 2^{n-2})$ matrix because the number of rows which $\bmath{X}$ has depends on elements of $\bmath{A}$ and $\bmath{B}$. For each $x_{\bullet,2j}(1\leq j\leq n-2)$ row in $\bmath{X}$, we create a 2 by 2 contingency table as shown in \textsf{\textbf{Table}} \textsf{\textbf{\ref{tb:tab2}}} and performed a chi-square binomial test. As $\bmath{B}$ contains all combination patterns for $\wedge$ and $\vee$ in its rows,  null hypothesis $H_0$ is that $\wedge$ and $\vee$ are equally likely to appear: both the number of $\wedge$ in $x_{\bullet,2j}$ and that of $\vee$ and $x_{\bullet,2j}$ are $N/2$ (\textbf{Supplemental material S.2}). When this chi-square binomial test has significant result and $\wedge (\vee)$ has higher proportion than the other, we define operator in $x_{\bullet,2j}$ has $\wedge (\vee)$ trend. As the order of columns in $\bmath{A}$ can change results, we have to consider an order of columns. We can use parentheses to change calculation oder. In this method, parentheses are used to define a variable from two or more existed variables.

In a dataset, in some cases, it is not practical to prepare $2^{n-2}\times n-2$ operator matrix $\bmath{B}$ because the number of rows increases exponentially. Instead of preparing $\bmath{B}$, we can allocate $\wedge$ and $\vee$ between explanatory variables by the same probability (1/2). Then we perform boolean calculations and check whether the allocated operators could explain an objective variable. If the operators can explain the objective variable, it is “faithful row,”  and if the operators cannot, it is unfaithful row. We can accumulate the faithful rows by repeating these steps and create 2 by 2 contingency table. Of course, it is not complete analyses compared with using $\bmath{B}$, however we can determine $\wedge(\vee)$ trend. 

\begin{table}
\small\sf\centering
\caption{Table created in the second step.}
\begin{tabular}{ccc} \toprule
\shortstack{For each $j$\\  $(1\leq j\leq n-2)$}&\shortstack{ The number of $\wedge$ \\ in $x_{\bullet,2j}$} & \shortstack{The number of $\vee$ \\ in $x_{\bullet,2j}$} \\ \midrule
Measured value& $\displaystyle\sum_{i=1}^Nf_{\wedge}(x_{i,2j})$ & $\displaystyle\sum_{i=1}^Nf_{\vee}(x_{i,2j})$ \\
Theoretical value& $\displaystyle\frac{N}{2}$ &  $\displaystyle\frac{N}{2}$ \\ 
\bottomrule
\label{tb:tab2}
\end{tabular}
\end{table}

\subsection*{Third step}

We assume that each operator in $x_{\bullet,2j}(1\leq j\leq n-2)$ has $\wedge(\vee)$ trend in the second step. Let $\bmath{O}$ be a $1\times n-2$ operator matrix. $o_{ij}$ which is the element of $\bmath{O}$ is defined as 
\[
  o_{1j}=\begin{cases}
    \wedge & (\textrm{when operator in } x_{\bullet,2j} \textrm{ has} \wedge \textrm{trend}) \\
    \vee & (\textrm{when operator in } x_{\bullet,2j} \textrm{ has} \vee \textrm{trend}).
  \end{cases}
\]
Then we calculate $m\times 2n-2$ dataset matrix $\bmath{D}=\bmath{A}\times \bmath{O}$. We create a 2 by 2 contingency table as shown in \textbf{\textsf{Table}} \textsf{\textbf{\ref{tb:tab3}}} and we can calculate some statistics such as sensitivity, specificity, positive predicting value and negative predicting value.

\begin{table}
\small\sf\centering
\caption{Table created in the third step.}
\begin{tabular}{ccc} \toprule
Property of rows & $d_{i,1}\cdots d_{i,2n-3}=1$ &$d_{i,1}\cdots d_{i,2n-3}=0$ \\ \midrule
$d_{i,2n-2}=1$& \#$_{\rm 1f}$($\bmath{D}$) & \#$_{\rm 1u}$($\bmath{D}$) \\
$d_{i,2n-2}=0$& \#$_{\rm 0u}$($\bmath{D}$) &  \#$_{\rm 1f}$($\bmath{D}$) \\ \bottomrule
\label{tb:tab3}
\end{tabular}
\end{table}

\setcounter{figure}{0}

\begin{figure}[htb]
\begin{center}
\includegraphics[width=7cm]{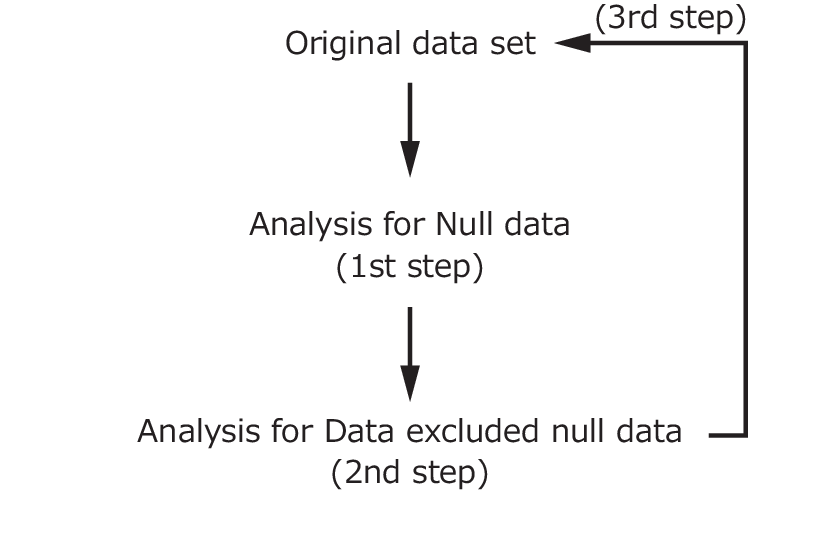}
\caption{The scheme of BMCM.  First step, null rows are analyzed. Second step, dataset matrix is analyzed and operators are determined. Third step, analysis for original dataset was performed using operators determined in second step. }
\label{fig:bmcm}
\end{center}
\end{figure}

In the third step, operators determined in the second step ($\bmath{O}$) are combined with $\bmath{A}$ and operations are performed in $\bmath{D}$. This step confirms how reliable the operators determined in the second step are. When one or more operators  do not have trend in the second step, we cannot perform the third step. We call this method ``Boolean Monte Carlo Method (BMCM)."  In \textsf{\textbf{Figure}} \textsf{\textbf{\ref{fig:bmcm}}}, the summary of three-step is shown.

All data were statistically analyzed using  Python Language Reference, version 3.6.2 (Python Software Foundation at {\tt http://www.python.org}). Statistical significance was defined as $P<0.05$.

\section{Example and Results}

In this sectoin, we applied the BMCM to a real dataset. This dataset can be obtained at website (\url{http://garthtarr.github.io/mplot/reference/diabetes.html}). This dataset contained ten baseline variables including age, body mass index (BMI) and mean arterial pressure (MAP) as well as the quantitative measure of disease progression (denoted by $Y$) one year after baseline, which were collected from 442 diabetes patients. Details of this dataset can be found elsewhere.\citep{efron2004}  We set cutoff to age, BMI, MAP and $Y$ and defined age$_B$, BMI$_B$, MAP$_B$ and $Y_B$ respectively. We regarded age$_B$, BMI$_B$, MAP$_B$ as explanatory variables and $Y_B$ as objective variable. Characteristics of patients  and definition of age$_B$, BMI$_B$, MAP$_B$ and $Y_B$ were shown in \textsf{\textbf{Table}} \textsf{\textbf{\ref{tb:tab4}}}. $\bmath{A}$ was a $4\times 442$ dataset matrix which were defined from age$_B$, BMI$_B$, MAP$_B$ and $Y_B$.  \textsf{\textbf{Table}} \textsf{\textbf{\ref{tb:summary}}} shows the number of each category. Positive likelihood ratio for age$_B$, BMI$_B$ and MAP$_B$ were 1.21, 2.00 and 1.65 respectively. As we considered three explanatory variables, there were three possible orders of explanatory variables; [Age$_B$, BMI$_B$, MAP$_B$], [Age$_B$, MAP$_B$, BMI$_B$] and [BMI$_B$, Age$_B$, MAP$_B$].  

\begin{table}
\small\sf\centering
\caption{Characteristics of patients.}
\begin{tabular}{lr} \toprule
Variables & Mean (SD)/ N($\%$) \\ \midrule
Age (years)& 48.5 (13.1)\\
\hspace{2ex}Age$\leq$ 45 (Age$_B=0_B$) & 165 (37.3\%) \\
\hspace{2ex}Age$\geq$ 45 (Age$_B=1_B$) & 277 (62.7\%) \\
BMI (kg/m$^2$)& 26.4 (4.4)\\
\hspace{2ex}BMI$\leq$ 25 (BMI$_B=0_B$) & 190 (43.3\%) \\
\hspace{2ex}BMI$\geq$ 25 (BMI$_B=1_B$) & 252 (56.7\%) \\
MAP (mmHg)& 94.6 (13.8) \\
\hspace{2ex}MAP$\leq$ 90 (MAP$_B=0_B$) & 192 (43.4\%) \\
\hspace{2ex}MAP$\geq$ 90 (MAP$_B=1_B$) & 250 (56.6\%) \\
$Y$& 152 (77) \\
\hspace{2ex}$Y\leq$ 210 ($Y_B=0_B$) & 331 (74.5\%) \\
\hspace{2ex}$Y\geq$ 210 ($Y_B=1_B$) & 111 (25.5\%) \\ \bottomrule
\end{tabular}
\\ \footnotesize{BMI, body mass index; MAP, mean artery pressure.}
\label{tb:tab4}
\end{table}

\begin{table}
\small\sf\centering
\caption{Category and the number of rows.}
\begin{tabular}{cccc|c}\hline
Age$_B$ & BMI$_B$ & MAP$_B$& $Y_B$   & N\\ \hline
$0_B$ & $0_B$ & $0_B$ & $0_B$ & 64 \\
$0_B$ & $0_B$ & $0_B$ & $1_B$ & 2  \\
$1_B$ & $0_B$ & $0_B$ & $0_B$ & 46 \\
$1_B$ & $0_B$ & $0_B$ & $1_B$ & 2  \\
$0_B$ & $1_B$ & $0_B$ & $0_B$ & 28 \\
$0_B$ & $1_B$ & $0_B$ & $1_B$ & 10 \\
$0_B$ & $0_B$ & $1_B$ & $0_B$ & 16 \\
$0_B$ & $0_B$ & $1_B$ & $1_B$ & 1  \\
$1_B$ & $1_B$ & $0_B$ & $0_B$ & 32 \\
$1_B$ & $1_B$ & $0_B$ & $1_B$ & 8  \\
$0_B$ & $1_B$ & $1_B$ & $0_B$ & 26 \\
$0_B$ & $1_B$ & $1_B$ & $1_B$ & 18 \\
$1_B$ & $0_B$ & $1_B$ & $0_B$ & 54 \\
$1_B$ & $0_B$ & $1_B$ & $1_B$ & 5  \\
$1_B$ & $1_B$ & $1_B$ & $0_B$ & 65 \\
$1_B$ & $1_B$ & $1_B$ & $1_B$ & 65 \\ \hline
\end{tabular}
\\ \footnotesize{BMI, body mass index; MAP,  \\ mean artery pressure, N; the number of rows.}
\label{tb:summary}
\end{table}

\subsection{First step}
$\bmath{A}$ was $442 \times 4$ boolean matrix.\textbf{\textsf{Table}} \textbf{\textsf{\ref{tb:tab6}}} shows the first 25 rows of  $\bmath{A}$. \textbf{\textsf{Table}} \textbf{\textsf{\ref{tb:tab11}}}  shows the contingency table for null data. Fisher's exact test showed that there was a significant difference in the proportions of $Y$ between the number of 1-rows  and that of 0-row $(\chi^2=40.9, p<0.0001,$ \textbf{\textsf{Table}} \textbf{\textsf{\ref{tb:tab11}}}).

\begin{table}
\small\sf\centering
\caption{First step. First 25 rows of $\bmath{A}$. Gray colored rows are null rows. 11 null rows were contained in the first 25 rows of $\bmath{A}$.}{\footnotesize
\begin{tabular}{c|cccc}
\hline 
$\bmath{A}$   & Age$_B$ & BMI$_B$ & MAP$_B$& $Y_B$ \\ \hline
$i/j$   & 1        & 2        & 3       & 4      \\ \hline
\rowcolor[rgb]{0.9, 0.9, 0.9}
1  & $1_B$        & $1_B$        & $1_B$       & $0_B$      \\
2  & $1_B$        & $0_B$        & $0_B$       & $0_B$      \\
\rowcolor[rgb]{0.9, 0.9, 0.9}
3  & $1_B$        & $1_B$        & $1_B$       & $0_B$      \\
4  & $0_B$        & $1_B$        & $0_B$       & $0_B$      \\
5  & $1_B$        & $0_B$        & $1_B$       & $0_B$      \\
\rowcolor[rgb]{0.9, 0.9, 0.9}
6  & $0_B$        & $0_B$        & $0_B$       & $0_B$      \\
\rowcolor[rgb]{0.9, 0.9, 0.9}
7  & $0_B$        & $0_B$        & $0_B$       & $0_B$      \\
\rowcolor[rgb]{0.9, 0.9, 0.9}
8  & $1_B$        & $1_B$        & $1_B$       & $0_B$      \\
9  & $1_B$        & $1_B$        & $0_B$       & $0_B$      \\
10 & $0_B$        & $1_B$        & $0_B$       & $1_B$      \\
11 & $0_B$        & $0_B$        & $1_B$       & $0_B$      \\
12 & $1_B$        & $1_B$        & $0_B$       & $0_B$      \\
13 & $1_B$        & $0_B$        & $1_B$       & $0_B$      \\
\rowcolor[rgb]{0.9, 0.9, 0.9}
14 & $1_B$        & $1_B$        & $1_B$       & $0_B$      \\
15& $1_B$        & $0_B$        & $1_B$       & $0_B$      \\
16 & $0_B$        & $0_B$        & $1_B$       & $0_B$      \\
\rowcolor[rgb]{0.9, 0.9, 0.9}
17 & $1_B$        & $1_B$        & $1_B$       & $0_B$      \\
\rowcolor[rgb]{0.9, 0.9, 0.9}
18 & $1_B$        & $1_B$        & $1_B$       & $0_B$      \\
19 & $0_B$        & $1_B$        & $0_B$       & $0_B$      \\
\rowcolor[rgb]{0.9, 0.9, 0.9}
20 & $0_B$        & $0_B$        & $0_B$       & $0_B$      \\
\rowcolor[rgb]{0.9, 0.9, 0.9}
21 & $0_B$        & $0_B$        & $0_B$       & $0_B$      \\
22 & $0_B$        & $0_B$        & $1_B$       & $0_B$      \\
23 & $0_B$        & $1_B$        & $1_B$       & $0_B$      \\
\rowcolor[rgb]{0.9, 0.9, 0.9}
24 & $1_B$        & $1_B$        & $1_B$       & $1_B$      \\
25 & $0_B$        & $1_B$        & $0_B$       & $0_B$      \\ \hline
\end{tabular}}
\\ \footnotesize{BMI, body mass index; MAP, mean artery pressure.}
\label{tb:tab6}
\end{table}

\begin{table}
\small\sf\centering
\caption{Null data analysis for $\bmath{A}$}
\begin{tabular}{lrr} \hline
$p<0.0001$& $Y_B=1$ & $Y_B=0$ \\ \hline
1-row &  64& 2 \\
0-row &  65& 65 \\ \hline
\end{tabular}
\label{tb:tab11}
\end{table}

\subsection{Second step}

We constructed boolean matrix  which had 2 columns because the number of explanatory variables is 3. As we randomly allocated 100 times 2 boolean operators and used only unique rows, it was almost equal to using  $4 \times 2$ boolean matrix $\bmath{B}$ shown in \textbf {Table} \textbf{\textsf{\ref{tb:tab7}}}. We show examples using $\bmath{B}$.
\textbf {Table} \textbf{\textsf{\ref{tb:tab8}}} shows first 56 rows of $\bmath{X}=\bmath{A}^-\otimes\bmath{B}$ when explanatory variables in $\bmath{A}$ has oder [Age$_B$, BMI$_B$, MAP$_B$].\textbf{\textsf{Table}} \textbf{\textsf{\ref{tb:faithful-count}}} shows first 24 rows of $\bmath{X}=(\bmath{A}^-\otimes\bmath{B})^{f}$ when explanatory variables in $\bmath{A}$ has oder [Age$_B$, BMI$_B$, MAP$_B$].

\begin{table}
\small\sf\centering
\caption{Boolean matrix $\bmath{B}$.}{%
\begin{tabular}{c|cc}
\hline
$i/j$   & 1        & 2        \\   \hline
1  & $\wedge$        & $\wedge$   \\
2  & $\wedge$        & $\vee$      \\
3  & $\vee$        & $\wedge$ \\
4  & $\vee$        & $\vee$      \\ \hline
\end{tabular}}
\label{tb:tab7}
\end{table}

\begin{table}
\small\sf\centering
\caption{Second step. First 56 rows of $\bmath{A}^-\otimes\bmath{B}$. Gray colored rows are 0-faithful row or 1-faithful row. The order of explanatory variables was [Age$_B$, BMI$_B$, MAP$_B$].}{\footnotesize
\begin{tabular}{c|cccccc}
\hline 
  $\bmath{A}^-\otimes\bmath{B}$&Age$_B$&&BMI$_B$&&MAP$_B$& $Y_B$ \\ \hline
  $i/j$ & 1        & 2        & 3        & 4        & 5       & 6      \\ \hline
\rowcolor[rgb]{0.9, 0.9, 0.9}
1  & $1_B$    & $\wedge$ & $0_B$    & $\wedge$ & $0_B$   & $0_B$  \\
\rowcolor[rgb]{0.9, 0.9, 0.9}
2  & $0_B$    & $\wedge$ & $1_B$    & $\wedge$ & $0_B$   & $0_B$  \\
\rowcolor[rgb]{0.9, 0.9, 0.9}
3  & $1_B$    & $\wedge$ & $0_B$    & $\wedge$ & $1_B$   & $0_B$  \\
\rowcolor[rgb]{0.9, 0.9, 0.9}
4  & $1_B$    & $\wedge$ & $1_B$    & $\wedge$ & $0_B$   & $0_B$  \\
5  & $0_B$    & $\wedge$ & $1_B$    & $\wedge$ & $0_B$   & $1_B$  \\
\rowcolor[rgb]{0.9, 0.9, 0.9}
6  & $0_B$    & $\wedge$ & $0_B$    & $\wedge$ & $1_B$   & $0_B$  \\
\rowcolor[rgb]{0.9, 0.9, 0.9}
7  & $1_B$    & $\wedge$ & $1_B$    & $\wedge$ & $0_B$   & $0_B$  \\
\rowcolor[rgb]{0.9, 0.9, 0.9}
8  & $1_B$    & $\wedge$ & $0_B$    & $\wedge$ & $1_B$   & $0_B$  \\
\rowcolor[rgb]{0.9, 0.9, 0.9}
9  & $1_B$    & $\wedge$ & $0_B$    & $\wedge$ & $1_B$   & $0_B$  \\
\rowcolor[rgb]{0.9, 0.9, 0.9}
10 & $0_B$    & $\wedge$ & $0_B$    & $\wedge$ & $1_B$   & $0_B$  \\
\rowcolor[rgb]{0.9, 0.9, 0.9}
11 & $0_B$    & $\wedge$ & $1_B$    & $\wedge$ & $0_B$   & $0_B$  \\
\rowcolor[rgb]{0.9, 0.9, 0.9}
12 & $0_B$    & $\wedge$ & $0_B$    & $\wedge$ & $1_B$   & $0_B$  \\
\rowcolor[rgb]{0.9, 0.9, 0.9}
13 & $0_B$    & $\wedge$ & $1_B$    & $\wedge$ & $1_B$   & $0_B$  \\
\rowcolor[rgb]{0.9, 0.9, 0.9}
14 & $0_B$    & $\wedge$ & $1_B$    & $\wedge$ & $0_B$   & $0_B$  \\
\rowcolor[rgb]{0.9, 0.9, 0.9}
15 & $1_B$    & $\wedge$ & $0_B$    & $\vee$   & $0_B$   & $0_B$  \\
\rowcolor[rgb]{0.9, 0.9, 0.9}
16 & $0_B$    & $\wedge$ & $1_B$    & $\vee$   & $0_B$   & $0_B$  \\
17 & $1_B$    & $\wedge$ & $0_B$    & $\vee$   & $1_B$   & $0_B$  \\
18 & $1_B$    & $\wedge$ & $1_B$    & $\vee$   & $0_B$   & $0_B$  \\
19 & $0_B$    & $\wedge$ & $1_B$    & $\vee$   & $0_B$   & $1_B$  \\
20 & $0_B$    & $\wedge$ & $0_B$    & $\vee$   & $1_B$   & $0_B$  \\
21 & $1_B$    & $\wedge$ & $1_B$    & $\vee$   & $0_B$   & $0_B$  \\
22 & $1_B$    & $\wedge$ & $0_B$    & $\vee$   & $1_B$   & $0_B$  \\
23 & $1_B$    & $\wedge$ & $0_B$    & $\vee$   & $1_B$   & $0_B$  \\
24 & $0_B$    & $\wedge$ & $0_B$    & $\vee$   & $1_B$   & $0_B$  \\
\rowcolor[rgb]{0.9, 0.9, 0.9}
25 & $0_B$    & $\wedge$ & $1_B$    & $\vee$   & $0_B$   & $0_B$  \\
26 & $0_B$    & $\wedge$ & $0_B$    & $\vee$   & $1_B$   & $0_B$  \\
27 & $0_B$    & $\wedge$ & $1_B$    & $\vee$   & $1_B$   & $0_B$  \\
\rowcolor[rgb]{0.9, 0.9, 0.9}
28 & $0_B$    & $\wedge$ & $1_B$    & $\vee$   & $0_B$   & $0_B$  \\
29 & $1_B$    & $\vee$   & $0_B$    & $\wedge$ & $0_B$   & $0_B$  \\
\rowcolor[rgb]{0.9, 0.9, 0.9}
30 & $0_B$    & $\vee$   & $1_B$    & $\wedge$ & $0_B$   & $0_B$  \\
31 & $1_B$    & $\vee$   & $0_B$    & $\wedge$ & $1_B$   & $0_B$  \\
32 & $1_B$    & $\vee$   & $1_B$    & $\wedge$ & $0_B$   & $0_B$  \\
33 & $0_B$    & $\vee$   & $1_B$    & $\wedge$ & $0_B$   & $1_B$  \\
\rowcolor[rgb]{0.9, 0.9, 0.9}
34 & $0_B$    & $\vee$   & $0_B$    & $\wedge$ & $1_B$   & $0_B$  \\
35 & $1_B$    & $\vee$   & $1_B$    & $\wedge$ & $0_B$   & $0_B$  \\
36 & $1_B$    & $\vee$   & $0_B$    & $\wedge$ & $1_B$   & $0_B$  \\
37 & $1_B$    & $\vee$   & $0_B$    & $\wedge$ & $1_B$   & $0_B$  \\
\rowcolor[rgb]{0.9, 0.9, 0.9}
38 & $0_B$    & $\vee$   & $0_B$    & $\wedge$ & $1_B$   & $0_B$  \\
\rowcolor[rgb]{0.9, 0.9, 0.9}
39 & $0_B$    & $\vee$   & $1_B$    & $\wedge$ & $0_B$   & $0_B$  \\
\rowcolor[rgb]{0.9, 0.9, 0.9}
40 & $0_B$    & $\vee$   & $0_B$    & $\wedge$ & $1_B$   & $0_B$  \\
41 & $0_B$    & $\vee$   & $1_B$    & $\wedge$ & $1_B$   & $0_B$  \\
\rowcolor[rgb]{0.9, 0.9, 0.9}
42 & $0_B$    & $\vee$   & $1_B$    & $\wedge$ & $0_B$   & $0_B$  \\
43 & $1_B$    & $\vee$   & $0_B$    & $\vee$   & $0_B$   & $0_B$  \\
44 & $0_B$    & $\vee$   & $1_B$    & $\vee$   & $0_B$   & $0_B$  \\
45 & $1_B$    & $\vee$   & $0_B$    & $\vee$   & $1_B$   & $0_B$  \\
46 & $1_B$    & $\vee$   & $1_B$    & $\vee$   & $0_B$   & $0_B$  \\
\rowcolor[rgb]{0.9, 0.9, 0.9}
47 & $0_B$    & $\vee$   & $1_B$    & $\vee$   & $0_B$   & $1_B$  \\
48 & $0_B$    & $\vee$   & $0_B$    & $\vee$   & $1_B$   & $0_B$  \\
49 & $1_B$    & $\vee$   & $1_B$    & $\vee$   & $0_B$   & $0_B$  \\
50 & $1_B$    & $\vee$   & $0_B$    & $\vee$   & $1_B$   & $0_B$  \\
51 & $1_B$    & $\vee$   & $0_B$    & $\vee$   & $1_B$   & $0_B$  \\
52 & $0_B$    & $\vee$   & $0_B$    & $\vee$   & $1_B$   & $0_B$  \\
53 & $0_B$    & $\vee$   & $1_B$    & $\vee$   & $0_B$   & $0_B$  \\
54 & $0_B$    & $\vee$   & $0_B$    & $\vee$   & $1_B$   & $0_B$  \\
55 & $0_B$    & $\vee$   & $1_B$    & $\vee$   & $1_B$   & $0_B$  \\
56 & $0_B$    & $\vee$   & $1_B$    & $\vee$   & $0_B$   & $0_B$ \\ \hline
\end{tabular}}
\\ \footnotesize{BMI, body mass index; MAP, mean artery pressure.}
\label{tb:tab8}
\end{table}

\begin{table}
\small\sf\centering
\caption{Second step. First 24 rows of $\bmath{X}=(\bmath{A}^-\otimes\bmath{B})^{f}$ and the number of $\wedge$ and $\vee$ in $x_{\bullet2}$ and $x_{\bullet4}$. }{\footnotesize
\begin{tabular}{c|cccccc}
\hline 
  $\bmath{X}=(\bmath{A}^-\otimes\bmath{B})^{f}$&Age$_B$&&BMI$_B$&&MAP$_B$& $Y_B$ \\ \hline
  $i/j$ & 1        & 2        & 3        & 4        & 5       & 6      \\ \hline
1  & $1_B$    & $\wedge$ & $0_B$    & $\wedge$ & $0_B$   & $0_B$  \\
2  & $0_B$    & $\wedge$ & $1_B$    & $\wedge$ & $0_B$   & $0_B$  \\
3  & $1_B$    & $\wedge$ & $0_B$    & $\wedge$ & $1_B$   & $0_B$  \\
4  & $1_B$    & $\wedge$ & $1_B$    & $\wedge$ & $0_B$   & $0_B$  \\
5  & $0_B$    & $\wedge$ & $0_B$    & $\wedge$ & $1_B$   & $0_B$  \\
6  & $1_B$    & $\wedge$ & $1_B$    & $\wedge$ & $0_B$   & $0_B$  \\
7  & $1_B$    & $\wedge$ & $0_B$    & $\wedge$ & $1_B$   & $0_B$  \\
8  & $1_B$    & $\wedge$ & $0_B$    & $\wedge$ & $1_B$   & $0_B$  \\
9 & $0_B$    & $\wedge$ & $0_B$    & $\wedge$ & $1_B$   & $0_B$  \\
10 & $0_B$    & $\wedge$ & $1_B$    & $\wedge$ & $0_B$   & $0_B$  \\
11 & $0_B$    & $\wedge$ & $0_B$    & $\wedge$ & $1_B$   & $0_B$  \\
12 & $0_B$    & $\wedge$ & $1_B$    & $\wedge$ & $1_B$   & $0_B$  \\
13 & $0_B$    & $\wedge$ & $1_B$    & $\wedge$ & $0_B$   & $0_B$  \\
14 & $1_B$    & $\wedge$ & $0_B$    & $\vee$   & $0_B$   & $0_B$  \\
15 & $0_B$    & $\wedge$ & $1_B$    & $\vee$   & $0_B$   & $0_B$  \\
16 & $0_B$    & $\wedge$ & $1_B$    & $\vee$   & $0_B$   & $0_B$  \\
17 & $0_B$    & $\wedge$ & $1_B$    & $\vee$   & $0_B$   & $0_B$  \\
18 & $0_B$    & $\vee$   & $1_B$    & $\wedge$ & $0_B$   & $0_B$  \\
19 & $0_B$    & $\vee$   & $0_B$    & $\wedge$ & $1_B$   & $0_B$  \\
20 & $0_B$    & $\vee$   & $0_B$    & $\wedge$ & $1_B$   & $0_B$  \\
21 & $0_B$    & $\vee$   & $1_B$    & $\wedge$ & $0_B$   & $0_B$  \\
22 & $0_B$    & $\vee$   & $0_B$    & $\wedge$ & $1_B$   & $0_B$  \\
23 & $0_B$    & $\vee$   & $1_B$    & $\wedge$ & $0_B$   & $0_B$  \\
24 & $0_B$    & $\vee$   & $1_B$    & $\vee$   & $0_B$   & $1_B$  \\ \hline
The number of $\wedge$&&17&&19& \\
The number of $\vee$&&7&&5& \\ \hline
\end{tabular}}
\\ \footnotesize{BMI, body mass index; MAP, mean artery pressure.}
\label{tb:faithful-count}
\end{table}

(i) A case the order of explanatory variables was  [Age$_B$, BMI$_B$, MAP$_B$]: In an equation Age$_B$\hspace{2pt}\fbox{1}\hspace{2pt}BMI$_B$\hspace{2pt}\fbox{2}\hspace{2pt}MAP$_B$$=Y_B$, we considered operators in \hspace{2pt}\fbox{1}\hspace{2pt} and \hspace{2pt}\fbox{2}\hspace{2pt}. A chi-square binomial test for faithful dataset matrix showed that operator in $x_{\bullet2}$ had $\wedge$ trend $(\chi^2=81.5, p<0.0001$,\textbf{\textsf{Table}} \textbf{\textsf{\ref{tb:tab12}}}) and  operator in $x_{\bullet4}$ had $\wedge$ trend $(\chi^2=38.8, p<0.0001$,\textbf{\textsf{Table}} \textbf{\textsf{\ref{tb:tab13}}}). This results determined \hspace{2pt}\fbox{1}\hspace{2pt}$=\wedge$ and \hspace{2pt}\fbox{2}\hspace{2pt}$=\wedge$.

\begin{table}
\small\sf\centering
\caption{In case (i), table for \hspace{2pt}\fbox{1}\hspace{2pt}.}
\begin{tabular}{ccc} \toprule
$p<0.0001$& Number of $\wedge$  in $x_{\bullet2}$ & Number of $\vee$  in $x_{\bullet2}$ \\ \midrule
Measured  value&  308& 121 \\
Theoretical  value& 214.5 &  214.5 \\ \bottomrule
\end{tabular}
\label{tb:tab12}
\end{table}

\begin{table}
\small\sf\centering
\caption{In case (i), table for \hspace{2pt}\fbox{2}\hspace{2pt}.}
\begin{tabular}{ccc} \toprule
$p<0.0001$& Number of $\wedge$  in $x_{\bullet4}$ & Number of $\vee$ in $x_{\bullet4}$ \\ \midrule
Measured value&  279& 150 \\
Theoretical value& 214.5 &  214.5 \\ \bottomrule
\end{tabular}
\label{tb:tab13}
\end{table}

(ii) A case the order of explanatory variables was  [Age$_B$, MAP$_B$, BMI$_B$]: In an equation,  Age$_B$\hspace{2pt}\fbox{1}\hspace{2pt}MAP$_B$\hspace{2pt}\fbox{2}\hspace{2pt}BMI$_B$, we considered operators in \hspace{2pt}\fbox{1}\hspace{2pt} and \hspace{2pt}\fbox{2}\hspace{2pt}. A chi-square binomial test for faithful dataset matrix showed that operator in $x_{\bullet2}$ had $\wedge$ trend $(\chi^2=79.5, p<0.0001$,\textbf{\textsf{Table}} \textbf{\textsf{\ref{tb:tab14}}}). and  operator in $x_{\bullet4}$ had $\wedge$ trend $(\chi^2=60.7, p<0.0001$,\textbf{\textsf{Table}} \textbf{\textsf{\ref{tb:tab15}}}). This results determined \hspace{2pt}\fbox{1}\hspace{2pt}$=\wedge$ and \hspace{2pt}\fbox{2}\hspace{2pt}$=\wedge$.

\begin{table}
\small\sf\centering
\caption{In case (ii), table for \hspace{2pt}\fbox{1}\hspace{2pt}.}
\begin{tabular}{ccc} \toprule
$p<0.0001$& Number of $\wedge$ in $x_{\bullet2}$ & Number of $\vee$  in $x_{\bullet2}$ \\ \midrule
Measured value&  305& 121 \\
Theoretical value& 213 &  213 \\ \bottomrule
\end{tabular}
\label{tb:tab14}
\end{table}

\begin{table}
\small\sf\centering
\caption{In case (ii), table for \hspace{2pt}\fbox{2}\hspace{2pt}.}
\begin{tabular}{ccc} \toprule
$p<0.0001$& Number of $\wedge$ in $x_{\bullet4}$ & Number of $\vee$  in $x_{\bullet4}$ \\ \midrule 
Measured  value&  279& 147 \\
Theoretical  value& 213 &  213 \\ \bottomrule
\end{tabular}
\label{tb:tab15}
\end{table}

(iii) A case the order of explanatory variables was  [BMI$_B$, Age$_B$, MAP$_B$]: In an equation BMI$_B$\hspace{2pt}\fbox{1}\hspace{2pt}Age$_B$\hspace{2pt}\fbox{2}\hspace{2pt}MAP$_B$$=Y_B$, we considered operators in \hspace{2pt}\fbox{1}\hspace{2pt} and \hspace{2pt}\fbox{2}\hspace{2pt}. A chi-square binomial test for faithful dataset matrix showed that operator in $x_{\bullet2}$ had $\wedge$ trend $(\chi^2=57.0, p<0.0001$,\textbf{\textsf{Table}} \textbf{\textsf{\ref{tb:tab16}}}). and  operator in $x_{\bullet4}$ had $\wedge$ trend $(\chi^2=52.8, p<0.0001$,\textbf{\textsf{Table}} \textbf{\textsf{\ref{tb:tab17}}}). This results determined \hspace{2pt}\fbox{1}\hspace{2pt}$=\wedge$ and \hspace{2pt}\fbox{2}\hspace{2pt}$=\wedge$.

\begin{table}
\small\sf\centering
\caption{In case (iii), table for \hspace{2pt}\fbox{1}\hspace{2pt}.}
\begin{tabular}{ccc} \toprule
$p<0.0001$& Number of $\wedge$ in $x_{\bullet2}$ & Number of $\vee$ in $x_{\bullet2}$ \\ \midrule 
Measured value&  308& 147 \\
Theoretical value& 227.5 &  227.5 \\ \bottomrule
\end{tabular}
\label{tb:tab16}
\end{table}

\begin{table}
\small\sf\centering
\caption{In case (iii), table for \hspace{2pt}\fbox{2}\hspace{2pt}.}
\begin{tabular}{ccc} \toprule
$p<0.0001$& Number of $\wedge$ in $x_{\bullet4}$ & Number of $\vee$  in $x_{\bullet4}$ \\ \midrule 
Measured value&  305& 150 \\
Theoretical value& 227.5 &  227.5 \\ \bottomrule
\end{tabular}
\label{tb:tab17}
\end{table}

\subsection{Third step}

In the third step, $\bmath{O}=[\wedge, \wedge]$ and $\bmath{D}=\bmath{A}\otimes\bmath{O}$.\textbf{\textsf{Table}} \textbf{\textsf{\ref{tb:tab18}}} shows the result using operators determined in the second step. Sensitivity, specificity, positive predicting value, negative predicting value and positive likelihood ratio were 0.50, 0.85. 0.59, 0.80 and 3.33 respectively.

\begin{table}
\small\sf\centering
\caption{Third step analysis.}
\begin{tabular}{ccc} \hline
& $d_{i1}d_{i2}d_{i3}d_{i4}d_{i5}=1$ &$d_{i1}d_{i2}d_{i3}d_{i4}d_{i5}=0$ \\ \hline
$d_{i6}=1$& 65 & 46 \\
$d_{i6}=0$& 65 &  266 \\ \hline
\end{tabular}
\label{tb:tab18}
\end{table}
\section{Discussion}
In this study, we proposed a new method (BMCM) in which we constructed 2 by 2 contingency table in a multivariate situation.
We applied this data to a real dataset which contains information of diabetes patients.  The American Diabetes Association (ADA) recommends that overweight adults with risk factors and all adults aged 45 years should be screened in using either an FPG test, A1C, or oral glucose tolerance test. The recommendations are partly based on substantial indirect evidence for the benefits of early treatment of type 2 diabetes and the fact that type 2 diabetes is typically present for years before clinical diagnosis.\cite{Kirkman2650}  Both diabetes and hypertension are cardiovascular risk factors, respectively and frequency increases with increasing age.\cite{FERRANNINI2012601} Guo et al.\cite{GUO2020} reported association mean arterial pressure and risk of type 2 diabetes. In the present study, we constructed 2 by 2 contingency tables setting a cutoff to age, BMI, MAP and the quantitative measure of disease progression. The result of this study indicate that  diabetes may be more progressive among a patient with age$\geq 45$, BMI$\geq 25$ and MAP$\geq90$. Of course, this result depend on a cutoff these variables. However, these results partly reconfirmed the validity of ADA recommendations. Positive likelihood ratios calculated for age$_B$, BMI$_B$ and MAP$_B$ by conventional ways were 1.21, 2.00 and 1.65 respectively whereas that calculated by BMCM was 3.33. When we simply multiply each  positive likelihood ratio in univariate case, we get $1.21\times 2.00\times 1.65\simeq 3.99$.  This difference might reflect interactions of variables. 

Applications of Boolean functions have been attempted in medicine. Previous studies applied boolean function mainly for the gene regulatory network (GRN). In these studies, to construct GRN models, operators between variables were determined by some methods such as the Bayesian approach\cite{lin_logic_2014,zhou_gene_2004}, Markov chain approach \cite{xiao_tutorial_2009}, and satisfiability problem solver (SAT solver) approach. \cite{Moskewicz_2001,lin_application_2012} These studies construct models of mutual interaction of genes by boolean operators. These mutual interactions often were interpreted as a GRN. Attempts to construct a GRN randomly were also reported. Kauffman \cite{kauffman_metabolic_1969} studied an approach for a GRN that was randomly connected. Pal R et al.\cite{pal_generating_2005} discussed a method to construct random attractors to examine GNR. As mentioned in method section, by assigning $\wedge$ and $\vee$ randomly, BMCM can be applied to a dataset of which the number of variables is large. To our knowledge, the random assignment of operators and the frequentist approach in determining operators have not been reported. We consider this method can be applied not only to genes but also to medical tests because the BMCM can construct a 2 by 2 contingency table in multivariate situation. This property of the method enables us to discuss the odds ratio in multivariate cases. In calculating odds ratio, logistic regression model is used for multivariate cases. Using a logistic regression model, we can calculate additive interactions of variables.\cite{hosmer_applied_2013} However, when we assume explanatory variables are independent with each other, we cannot consider non-additive interactions. Pepe et al.\cite{pepe_limitations_2004} pointed out that there was a pitfall in using the logistic regression model for medical markers. They argued that strong associations are required for meaningful classification accuracy in using the logistic regression model. The BMCM has a disadvantage in weighting variables, whereas it may have an advantage in modeling interactions because operator can be interpreted as interaction. To clarify property of  BMCM, further study should be done.
This study has some limitations. First, this method can be applied only to binomial data and variables cannot be weighted. However, this model can determine the relations of variables. BMCM can be an option in analysis. Second, an interpretation of results can be complex because the number of orders of variables increase exponentially. However, we can estimate the order of variables from previous studies. Third, when a 2 by 2 contingency table is written, we can use different functions. For example, we can choose a function to maximize sensitivity whereas we can choose another function to maximize specificity. There can be many cross tables. However, in this case, we can use each table case by case.

\section{Conclusion}

We introduced boolean Monte Carlo method (BMCM) in which we can construct a 2 by 2 contingency table to multivariate situation by frequentist approach. We may be able to detemine trends of boolean operators and construct a 2 by 2 contingency table in multivariate situation by this method.

\section{Conflict of Interest}

All authors have no conflict of interest to disclose with respect to this research.

\begin{sm}
\subsection*{S.1. Python code}

%\lstinputlisting[language=Python]{code_paper.py}
\begin{lstlisting}[language=Python]
import numpy as np
import random
import pandas as pd

"""

Boolean matrix A.
A = pd.DataFrame(np.array([
        [0, 0, 0, 0],
        [0, 0, 0, 1],
        [1, 0, 0, 1],
        [0, 1, 0, 1],
        [0, 0, 1, 1]
        ]))

Operator matrix B
B = pd.DataFrame(np.array([
        ['and', 'and'],
        ['and', 'or'],
        ['or', 'and'],
        ['or', 'or']
        ]))

Using "null_rows_deriver", we can derive all null rows from boolean matrix A.
We need to input column number of objective variabls.

Example:
null_rows_deriver(A, 3)

Result:
   0  1  2  3
1  0  0  0  0
2  0  0  0  1
(Notice that an index starts from 0 in Python.)

Using "null_data_deleter", you can obtain a matrix (A^-)
which does not contain any null row.
We need to input column number of objective variabls.

Example:
null_rows_deleter(A, 3)

Result:
   0  1  2  3
2  1  0  0  1
3  0  1  0  1
4  0  0  1  1

(Notice that an index starts from 0 in Python.)

"""


def null_rows_deriver(matrix, objective):
    m, n = matrix.shape
    matrix_variable = matrix.drop(matrix.columns[objective], axis=1)
    index_list = []
    for i in range(m):
        value_list = [x for x in matrix_variable.iloc[i, :]]
        if len(set(value_list)) == 1:
            if 0 in value_list:
                index_list.append(i)
            elif 1 in value_list:
                index_list.append(i)
    matrix_answer = matrix.iloc[index_list, :]
    return(matrix_answer)


def null_rows_deleter(matrix, objective):
    m, n = matrix.shape
    matrix_variable = matrix.drop(matrix.columns[objective], axis=1)
    index_list = []
    for i in range(m):
        value_list = [x for x in matrix_variable.iloc[i, :]]
        if len(set(value_list)) == 1:
            if 0 in value_list:
                index_list.append(i)
            elif 1 in value_list:
                index_list.append(i)
    matrix_answer = matrix.drop(matrix.index[index_list])
    return(matrix_answer)


"""

If we have full boolean matrix "B",
we can use function "otimes" to construct "X"
from A^- and B.

Exaple:
A_minus = null_rows_deleter(A, 3)
print(otimes(A_minus, B))

Result:
    0    1  2    3  4  5
0   1  and  0  and  0  1
1   1  and  0   or  0  1
2   1   or  0  and  0  1
3   1   or  0   or  0  1
4   0  and  1  and  0  1
5   0  and  1   or  0  1
6   0   or  1  and  0  1
7   0   or  1   or  0  1
8   0  and  0  and  1  1
9   0  and  0   or  1  1
10  0   or  0  and  1  1
11  0   or  0   or  1  1

(Notice that an index starts from 0 in Python.)

"""


def otimes(boolean_matrix, operator_matrix):
    a = boolean_matrix
    b = operator_matrix
    m, n = a.shape
    l, k = b.shape
    c = pd.DataFrame(index=["{}".format(i) for i in range(m*l)],
                     columns=["{}".format(i) for i in range(2*n-2)])
    if k != n-2:
        print('Error! You should check the number of columns in B')
        return(0)
    else:
        for i in range(m*l):
            for j in range(2*n-2):
                if j % 2 == 0 and j < 2*n-3:
                    c.iloc[i, j] = a.iloc[(i//l), int(j/2)]
                if j % 2 == 1 and j < 2*n-3:
                    c.iloc[i, j] = b.iloc[(i % l), int((j-1)/2)]
                if j == 2*n-3:
                    c.iloc[i, j] = a.iloc[(i//l), n-1]
        return(c)


"""

When we cannot obtain full boolean matrix,
we can use function "random_operato_assignment".
Using this function, we can get a part of X.
We can input the number of allocation.

Example:
A_minus = null_rows_deleter(A, 3)
print(random_operato_assignment(A_minus, 100))

Result:
    0    1  2    3  4  5
0   1  and  0  and  0  1
1   0  and  1   or  0  1
2   0  and  0  and  1  1
3   1   or  0   or  0  1
4   0   or  1   or  0  1
5   1   or  0  and  0  1
6   0  and  1  and  0  1
7   1  and  0   or  0  1
8   0   or  0   or  1  1
9   0   or  1  and  0  1
10  0  and  0   or  1  1
11  0   or  0  and  1  1

(Notice that an index starts from 0 in Python.)

"""


def random_operato_assignment(boolean_matrix, N):
    A = boolean_matrix
    m, n = A.shape
    varlistA = [x for x in A.columns]
    varlistO = ['O{}'.format(i) for i in range(n-2)]
    for i in range(n-2):
        A['O{}'.format(i)] = 0
    varlist = [varlistA[int(i/2)] if i % 2 == 0 else varlistO[int((i-1)/2)]
               for i in range(2*n-3)]
    varlist.append(varlistA[-1])
    A = A[varlist]
    A_app = A
    for i in range(N-1):
        A = pd.concat([A, A_app])
    for j in range(n-2):
        random_and_or = ['and' if random.randint(0, 1) == 1
                         else 'or' for k in range(m*N)]
        A.iloc[:, 2*j+1] = random_and_or
    A = A.drop_duplicates()
    A = A.reset_index(drop=True)
    A.columns = [x for x in range(2*n-2)]
    return(A)


"""

Using the function "logical_operation_executor",
we can obtain whether a row in X is faithful or not.

Example:
A_minus = null_rows_deleter(A, 3)
X = otimes(A_minus, B)
logical_operation_executor(X)

Results:
0 th row is unfaithful row.
1 th row is unfaithful row.
2 th row is unfaithful row.
3 th row is faithful row.
4 th row is faithful row.
5 th row is faithful row.
6 th row is unfaithful row.
7 th row is unfaithful row.
8 th row is faithful row.
9 th row is unfaithful row.
10 th row is faithful row.
11 th row is unfaithful row.

"""


def logical_operation_executor(dataset_matrix):
    X = dataset_matrix
    m, n = X.shape
    for i in range(m):
        logical_equation = ''
        for j in range(n):
            if j % 2 == 0 and j == 0:
                logical_equation = logical_equation + \
                    '({} '.format(X.iloc[i, j])
            elif j % 2 == 0 and 0 < j < n-1:
                logical_equation = logical_equation + \
                    ' {} '.format(X.iloc[i, j])
            elif j % 2 == 1 and j < n-1:
                logical_equation = logical_equation + X.iloc[i, j]
            elif j == n-1:
                logical_equation = logical_equation +\
                    ') == {}'.format(X.iloc[i, j])
        if eval(logical_equation, globals()) is True:
            print('{} th row is faithful row.'.format(i))
        elif eval(logical_equation, globals()) is False:
            print('{} th row is unfaithful row.'.format(i))
    return(0)

\end{lstlisting}

\subsection*{S.2. Mathematical Background}

We review mathematical background of BMCM. We refer to \cite{hogg_introduction_2019} for discussion.
$b(n, p)$ denotes binomial distribution with probability $p$ and degree of freedom $n$.
Let $X$ be $b(n, p)$ and we consider the random variable
\begin{equation*}
Y=\frac{X-np}{\sqrt{np(1-p)}}
\end{equation*}
which has, as $n \rightarrow \infty$, an approximate $N(0, 1)$ distribution (Central limit theorem). Furthermore, $Y^2$ is approximately $\chi^2(1)$.
\begin{equation*}
Y^2=\frac{(X-np)^2}{np(1-p)}=\frac{(X-np)^2}{np}+\frac{(X-np)^2}{n(1-p)}\sim \chi^2(1)
\end{equation*}
Chi square test is based on this property of distribution.  For instance, we explain how to determine operator by chi-square test. We use the case the order of variables is [Age, BMI, MAP] in the \textbf{Example and Results} section.  In chi square test for BMCM, the hypothesis $H_0$: both the number of $\wedge$ in $x_{\bullet,2j}$ and that of $\vee$ and $x_{\bullet,2j}$ are $N/2$: $p=1/2$ was tested. We demonstrate determination process of \hspace{2pt}\fbox{1}\hspace{2pt} in  Age$_B$\hspace{2pt}\fbox{1}\hspace{2pt}BMI$_B$\hspace{2pt}\fbox{2}\hspace{2pt}MAP$_B$$=Y_B$. In faithful dataset matrix, the frequency of  $\wedge$  in  $x_{\bullet,2}$ was 308 and that of  $\vee$ was 121. As operators were randomly assigned by the same probability $1/2$,
\begin{equation*}
Y^2=\frac{\left(308-429\times \frac12\right)^2}{429\times \frac12}+\frac{\left(121-429\times \frac12\right)^2}{429\times \left(1-\frac12\right)}\simeq 81.5
\end{equation*}
and determine \hspace{2pt}\fbox{1}\hspace{2pt}.

\end{sm}

\newpage
\bibliographystyle{SageV}
\bibliography{SMMR20200406}
\end{document}